\documentclass[singlecolumn,prl,12pt,nofootinbib]{revtex4}
\usepackage{graphicx}
\usepackage{dcolumn}
\usepackage{amsmath,amsthm,amssymb}

 \newcommand{\arXiv}[1]{\href{http://www.arXiv.org/abs/#1}{#1}}
\usepackage[colorlinks=true, linkcolor=blue, bookmarks=true]{hyperref}

\theoremstyle{remark}

\begin{document}

\title{Gravitational turbulent instability of AdS$_5$}
\author{Piotr Bizo\'n}
\affiliation{Institute of Physics, Jagiellonian University, Krak\'ow, Poland}
\author{Andrzej Rostworowski}
\affiliation{Institute of Physics, Jagiellonian University, Krak\'ow, Poland}
%
%
\begin{abstract}
  We consider the problem of stability of anti-de Sitter spacetime in five dimensions under small purely gravitational  perturbations satisfying the cohomogeneity-two biaxial Bianchi IX ansatz. In analogy to spherically symmetric scalar perturbations, first studied in \cite{br, jrb}, we observe numerically a black hole formation on the time-scale $\mathcal{O}(\varepsilon^{-2})$, where $\varepsilon$ is the size of the perturbation.
\end{abstract}
\maketitle
\section{Introduction}
Over the past nearly two decades asymptotically anti-de Sitter (aAdS) spacetimes have
received a great deal of attention, primarily due to the AdS/CFT
correspondence which is the conjectured duality between aAdS  spacetimes and conformal field theories.
 The distinctive  feature of aAdS spacetimes, on which the very concept of  duality  rests,
is a timelike conformal boundary at spatial and null infinity, where it is necessary to specify boundary conditions in order to define the deterministic evolution. For  energy conserving boundary conditions the conformal boundary acts as a mirror at which  massless waves propagating outwards bounce off and return to the bulk.  Therefore, the key mechanism stabilizing the evolution of asymptotically flat spacetimes -- dispersion of energy by radiation -- is absent in aAdS spacetimes. For this reason  the problem of
nonlinear stability of the pure AdS spacetime   (which is the ground state among aAdS spacetimes) is particularly  challenging.

Two few  years ago we considered this problem in a toy model of the spherically symmetric massless scalar field minimally coupled to  gravity with a negative cosmological constant  in  four and higher dimensions and  gave evidence for the instability of the AdS spacetime \cite{br,jrb}. More precisely, we showed numerically that there is a large class of arbitrarily small
perturbations of AdS that evolve into a black hole on the time-scale $\mathcal{O}(\varepsilon^{-2})$, where $\varepsilon$
is the size of the perturbation. On the basis of nonlinear perturbation analysis, we
conjectured that this instability is due to a resonant transfer
of energy from low to high frequencies, or equivalently, from coarse to fine spatial
scales, until eventually an apparent horizon forms.

  Further studies of this and similar models confirmed and extended our findings and provided important new insights concerning the  coexistence of unstable (turbulent) and stable (quasiperiodic) regimes of evolution (see \cite{ce} for a brief review and references).

  The major downside of all the reported numerical simulations of AdS instability was the restriction to spherical symmetry, so that no gravitational degrees of freedom were excited.
   Nonetheless, the results of nonlinear perturbation analysis of the vacuum Einstein equation without symmetry assumptions \cite{dhs} seem to indicate that the instability mechanism is present for gravitational perturbations as well. To verify this expectation,
  in this paper we consider the vacuum Einstein equations with negative cosmological constant in five dimensions within the cohomogeneity-two biaxial Bianchi IX ansatz, introduced originally in the context of critical collapse for asymptotically flat spacetimes \cite{bcs}. This ansatz provides a simple 1+1 dimensional setting for analyzing  stability  of AdS$_5$ (which, incidentally, happens to be dimension\-wise the most interesting case from the AdS/CFT viewpoint) under purely gravitational perturbations. As expected, we observe a similar instability phenomenon as in \cite{br,jrb}. In the remaining two sections we describe the model and  present numerical results.
\section{Setup} We consider the vacuum Einstein equations with negative cosmological constant in five dimensions
\begin{equation}\label{eve}
  R_{\alpha\beta}=-\frac{4}{\ell^2}\, g_{\alpha\beta}.
\end{equation}
Following \cite{bcs}, we assume
 the cohomogeneity-two biaxial Bianchi IX ansatz
\begin{equation}
\label{bcs}
g = \frac{\ell^2}{\cos^2{\!x}}\left( -A e^{-2 \delta} dt^2 + A^{-1} dx^2
+ \frac{1}{4} \sin^2{\!x} \left(e^{-2B} (\sigma_1^2+\sigma_2^2)+e^{4B} \sigma_3^2\right)\right)\,,
\end{equation}
where $(t,x)\in \mathbb{R}\times [0,\pi/2)$ and $A$, $\delta$, $B$ are functions of $(t,x)$. The angular part of this metric is the $SU(2)\times U(1)$-invariant homogeneous metric on the squashed 3-sphere. Here $\sigma_k$ are left-invariant one-forms on $SU(2)$ which in terms of Euler angles  ($0\leq\vartheta\leq \pi, 0\leq \varphi,\psi\leq 2\pi$) take the form
\begin{equation}
\sigma_1+i\sigma_2=e^{i\psi}(\cos{\vartheta} d\varphi+i d\vartheta),\quad \sigma_3=d\psi-\sin{\vartheta} d\varphi\,.
\end{equation}
Note that if $B=0$ the angular metric becomes the round metric on $S^3$ and the symmetry is enhanced to $SO(4)$.

  For the ansatz \eqref{bcs} the Einstein equations \eqref{eve} reduce to the following  1+1 dimensional system (hereafter, primes and overdots denote derivatives with respect to $x$ and $t$, respectively)
\begin{align}
\label{wave}
\dot B &= A e^{-\delta} P, \qquad \dot P = \frac{1}{\tan^3{\!x}} \left(\tan^3{\!x}\, A e^{-\delta} Q \right)'-\frac{4 e^{-\delta}}{3\sin^2{\!x}}\left(e^{-2B}-e^{-8B}\right),\\
\label{ap}
A' &= 4 \tan{x} \, (1-A) -  2\sin{x} \cos{x} \, A \left(Q^2 + P^2 \right)
+\frac{2(4e^{-2B}-e^{-8B}-3A)}{3\tan{x}}\,,
\\
\label{dp}
\delta' &= -2\sin{x} \cos{x} \left(Q^2+P^2\right) \,,
\\
\label{ad}
\dot A &= - 4 \sin{x} \cos{x} \, A^2 e^{-\delta} Q P \,,
\end{align}
where we have introduced the auxiliary variables $Q=B'$ and $P=A^{-1} e^{\delta} \dot B$.
The field $B$ is the only dynamical degree of freedom which plays a role similar to the spherical scalar field\footnote{If $B=0$, the only solution is the Schwarzschild-AdS family, in agreement with the Birkhoff theorem.}. It is convenient to define the mass function
\begin{equation}\label{mass-function}
m(t,x) = \frac{\sin^2{x}}{\cos^4{x}}\,(1-A(t,x)).
\end{equation}
From the Hamiltonian constraint (\ref{ap}) it follows that
\begin{equation}
m'(t,x) = 2\left[A(Q^2+P^2) + \frac{1}{3\sin^2 x} \left( 3 + e^{-8B} - 4e^{-2B} \right)\right] \tan^3 x\geq 0\,.
\end{equation}
We want to solve the system (4-6) for small smooth initial data with finite total mass $M = \lim_{x\rightarrow\pi/2} m(t,x)$. Smoothness at $x=0$ implies that
\begin{equation}\label{x=0}
   B(t,x)= b_0(t)\,x^2+\mathcal{O}(x^4), \quad\delta(t,x)= \mathcal{O}(x^4),\quad
    A(t,x)=1+\mathcal{O}(x^4),
\end{equation}
where we used normalization $\delta(t,0)=0$ to ensure  that $t$ is the proper time at the origin.
The power series  \eqref{x=0} are uniquely determined by the free function $b_0(t)$.
Smoothness at $x=\pi/2$ and finiteness of the total mass $M$ imply that  (using $\rho=x-\pi/2$)
\begin{equation}\label{pi2}
     B(t,x)= b_{\infty}(t)\, \rho^4+\mathcal{O}\left(\rho^6\right),\quad
    \delta(t,x)= \delta_{\infty}(t)+\mathcal{O}\left(\rho^8\right),\quad
    A(t,x)= 1-M \rho^4+\mathcal{O}\left(\rho^6\right)\,,
\end{equation}
where the free functions $b_{\infty}(t)$, $\delta_{\infty}(t)$, and  mass $M$ uniquely determine the power series.
  It follows from \eqref{pi2} that the asymptotic behaviour of fields at  infinity is completely fixed by the  assumptions of smoothness and finiteness of total mass, hence there is no freedom of imposing the boundary data.

  The pure AdS spacetime corresponds to $B=0,A=1,\delta=0$. Linearizing around this solution, we obtain
  \begin{equation}\label{L}
     \ddot B +L B=0 ,\qquad
    L=-\frac{1}{\tan^3{\!x}}\, \partial_x \left(\tan^3{\!x} \,\partial_x\right)+\frac{8}{\sin^2{\!x}}\,.
\end{equation}
This equation is the $\ell=2$ gravitational tensor case of the  master equation describing the evolution of linearized perturbations of AdS spacetime, analyzed in detail by Ishibashi and Wald \cite{iw}.
The Sturm-Liouville operator $L$ is essentially self-adjoint with respect to the inner product $(f,g)=\int_0^{\pi/2} f(x) g(x) \tan^3{\!x}\,dx$. The eigenvalues and  associated orthonormal eigenfunctions of $L$ are ($k=0,1,\dots$)
\begin{equation}\label{modes}
\omega^2_k=(6+2k)^2,\qquad e_k(x)= 2 \sqrt{\frac{(k+3)(k+4)(k+5)}{(k+1)(k+2)}}\, \sin^2{\!x} \cos^4{\!x} \,P_k^{(3,2)}(\cos{2x})\,,
\end{equation}
where $P_k^{(a,b)}(x)$ is a Jacobi polynomial of order $k$.

The eigenfunctions $e_k(x)$ fulfill the regularity conditions \eqref{x=0} and \eqref{pi2} hence any smooth solution can be expressed as
\begin{equation}
B(t,x)=\sum\limits_{k\geq 0} b_k(t) e_k(x)\,.
\end{equation}
To quantify the transfer of energy between the modes we introduce the linearized energy
\begin{equation}\label{E}
E = \int_0^{\pi/2} \left( \dot B^2 + B'^2 + \frac{8}{\sin^2 x} B^2 \right) \tan^3 x\, dx=\sum\limits_{k\geq 0} E_k,
\end{equation}
where $E_k=\dot b_k^2 + \omega_k^2 b_k^2$ is the energy of the $k$-th mode.
\section{Numerical results}
We solve the system \eqref{wave}-\eqref{dp} numerically for small smooth initial data and the boundary conditions \eqref{x=0} and \eqref{pi2}. We use the standard method of lines with a fourth-order Runge-Kutta time integration and fourth-order spatial finite differences.  Kreiss-Oliger dissipation is added  to eliminate high-frequency instabilities. The scheme is fully constrained, that is the metric functions $A$ and $\delta$ are updated at each time step by solving the hamiltonian constraint \eqref{ap} and the slicing condition \eqref{dp}. The momentum constraint \eqref{ad} is only monitored to verify the accuracy of computations.
For typical initial data the energy is rapidly transferred to small spatial scales. To resolve these scales, we refine the entire spatial grid when a global spatial error exceeds some prescribed tolerance level. We usually start on a grid with $2^{14}+1$ points and allow for four levels of refinement. To feel confident that the spatial scales are properly resolved, we validated our computations by convergence tests.

  The results presented below were generated from the Gaussian initial data of the form
  \begin{equation}\label{data}
  B(0,x)=0\,,\quad P(0,x)=\varepsilon \left(\frac 2 \pi \right)^3 \frac{512}{\sqrt{3}} x^2\exp(- 4 \tan^2{\!x}/ (\pi^2 \sigma^2))
  \end{equation}
   with width $\sigma=1/16$ and varying small amplitudes $\varepsilon$. A good  indicator of instability is the Kretschmann scalar at the origin
    \begin{equation}\label{kret}
 R_{\alpha\beta\gamma\delta} R^{\alpha\beta\gamma\delta}(t,0)=40+864 \,Q'(t,0)^2\,.
 \end{equation}
For initially small  narrow wave packets such as \eqref{data}, this quantity oscillates  with a period nearly equal to $\pi$, while the amplitude of oscillations grows exponentially. As shown in Fig.~1, the one-period maxima of the quantity $\varepsilon^{-2} Q'(\varepsilon^2 t,0)$
 are almost independent of $\varepsilon$.
  \begin{figure}[h]
 \includegraphics[width=0.48\textwidth]{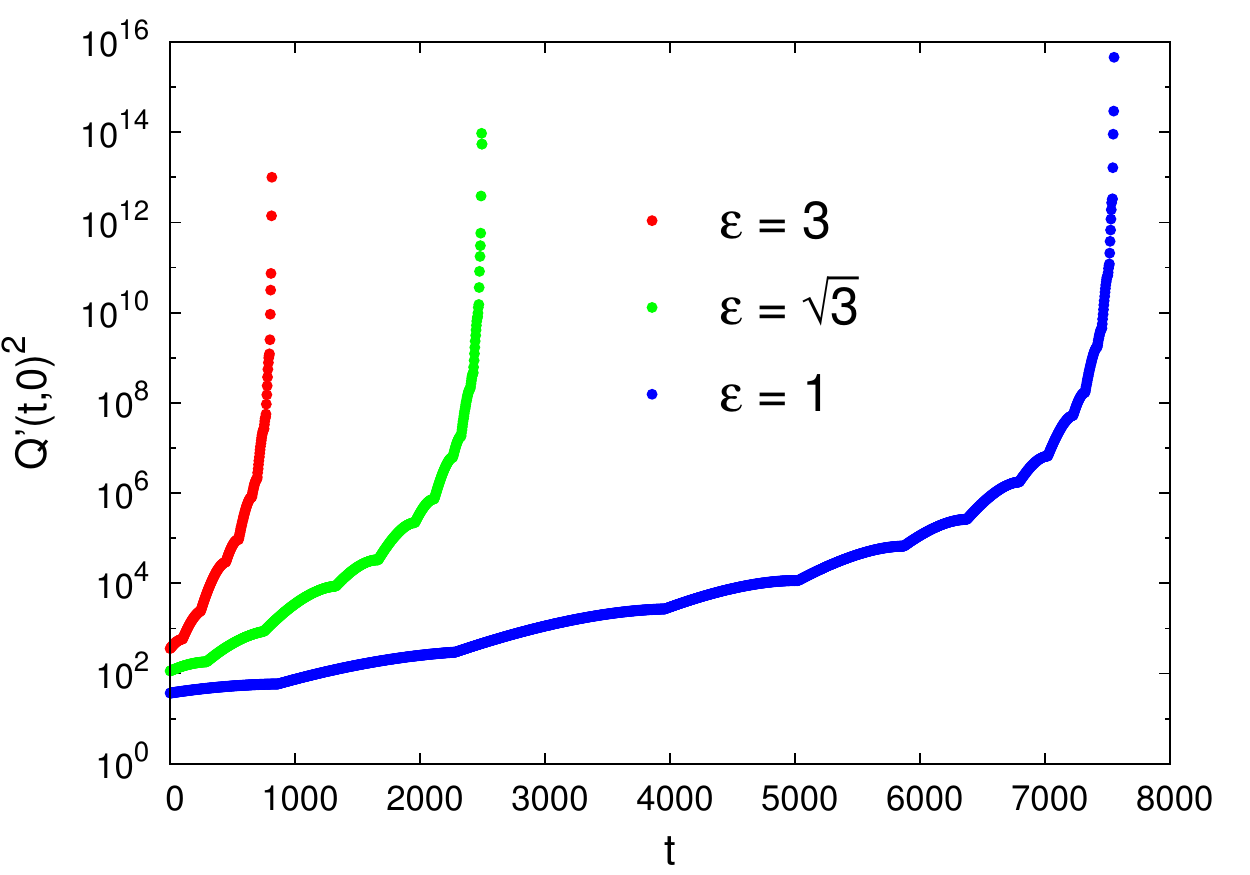}
 \includegraphics[width=0.48\textwidth]{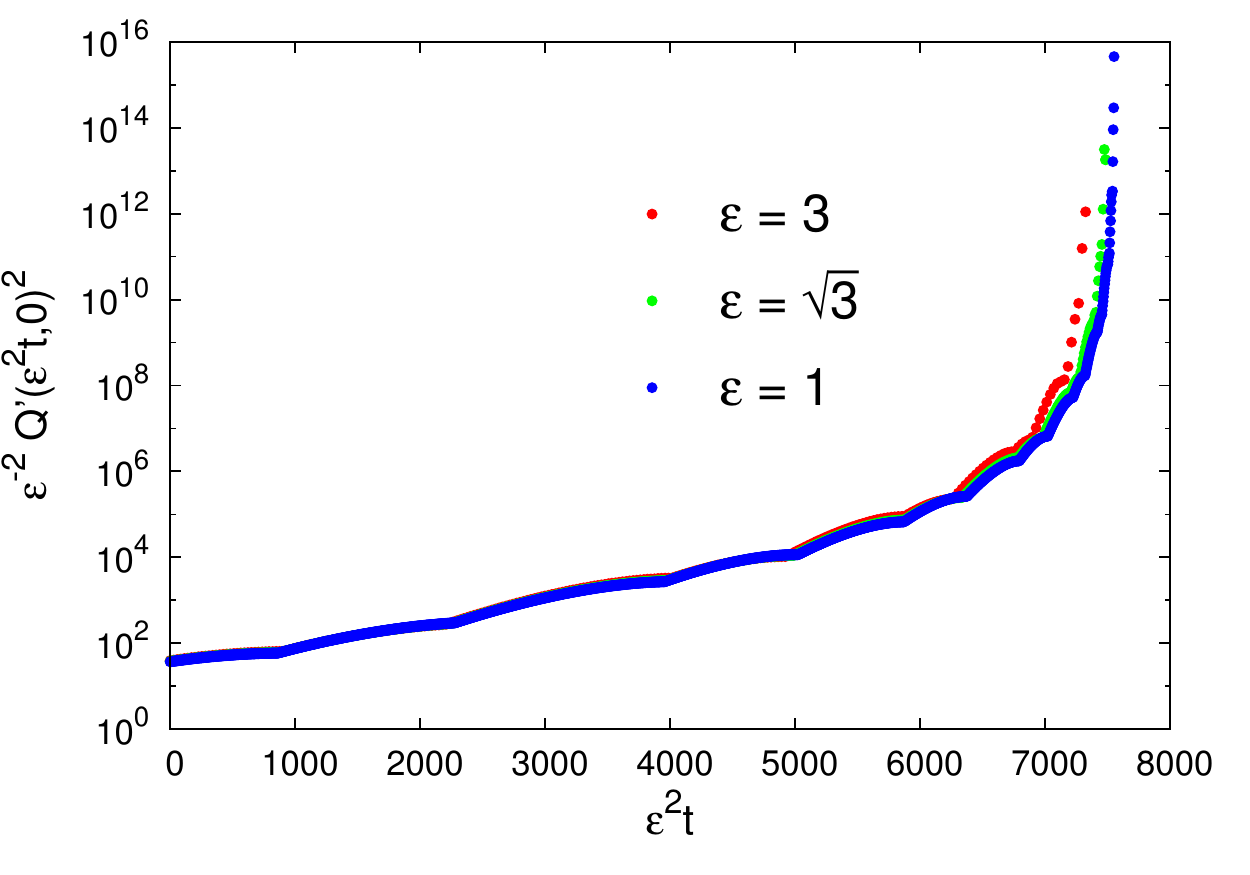}
  \caption{Left: One-period maxima of $Q'(t,0)^2$ for the initial data \eqref{data} with three relatively small amplitudes (note, however, that for these data $Q'(0,0)^2$ is much bigger than the unperturbed Kretschmann scalar at the origin). Right: After rescaling $\varepsilon^{-2} Q'(\varepsilon^2 t)^2$, the curves from the left panel  nearly coincide.}
 \label{spectrum}
\end{figure}

 At the end of evolutions shown in Fig.~1 we observe  the formation of an apparent horizon which is signalled by $A(t,r)$ dropping below a certain small threshold (we take this threshold to be $2^{-7}$ on the $N=2^{14}$ grid and then divide it by two each time we increase the grid resloution by a factor two).
 Since the computational costs of numerical simulations rapidly increase with decreasing $\varepsilon$, we have not been able to determine the outcome of evolution of smaller perturbations. Nonetheless, extrapolating the observed scaling behavior, we conjecture that  AdS$_5$ is unstable against black hole formation for a large class of arbitrarily small purely gravitational perturbations.

 On a heuristic level the mechanism of instability is the same as for scalar perturbations, namely the turbulent cascade of energy from low to high modes that is eventually cut-off by the formation of a black hole\footnote{We stress that this instability is not active for some perturbations. In particular, there exist initial data for which the solutions are exactly time-periodic \cite{m}.}. To substantiate this claim, let us see how the energy of the perturbation gets distributed over the modes in the course of evolution. To this end, in Fig.~2 we depict the linear energy spectrum, as defined in \eqref{E}.
 The range of modes participating in the evolution is seen to increase very rapidly. Just before collapse the spectrum exhibits the power-law scaling $E \sim k^{-\alpha}$, where the exponent $\alpha\approx 5/3$ appears to be universal, i.e., independent of initial data (for comparison, in the Einstein-scalar AdS collapse in five dimensions $\alpha\approx 2$ \cite{bmr}). This power-law spectrum reflects the loss od smoothness of the collapsing solution.
 \begin{figure}[h]
 \includegraphics[width=0.7\textwidth]{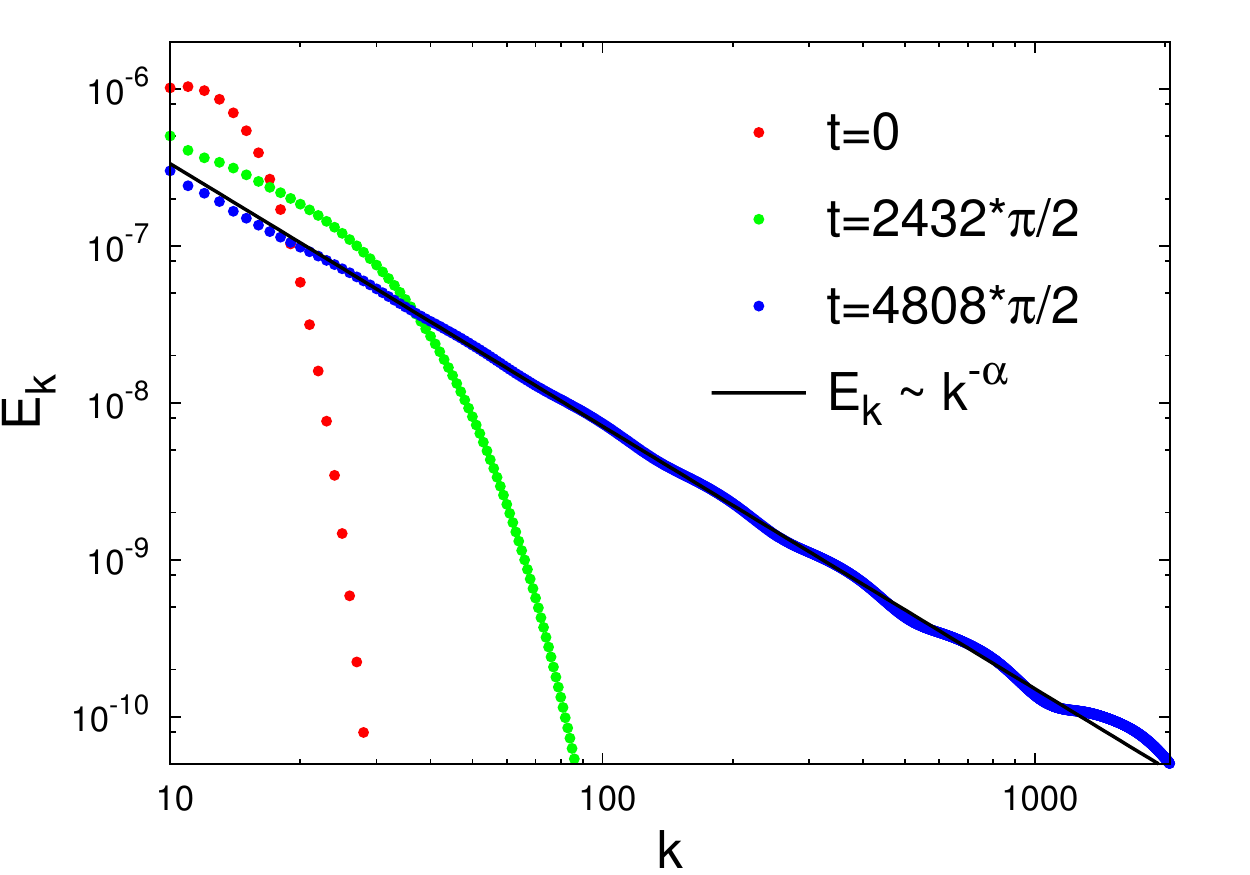}
  \caption{Log-log plot of the linear energy spectra at three instants of time for the initial data \eqref{data} with $\varepsilon=0.3$. The fit  at $t \approx 4808 \pi /2$ yields a power-law spectrum $E_k \sim k^{-\alpha}$ with $\alpha \approx 5/3$.}
 \label{spectrum}
\end{figure}

 \section{Acknowledgments}

This work   was supported by
the Polish National Science Centre grant no. DEC-2012/06/A/ST2/00397.

\end{document}